\documentclass[%
reprint, twocolumn,
superscriptaddress,
showpacs,preprintnumbers,
nofootinbib,
amsmath,amssymb, aps, pra,
]{revtex4}

\usepackage{graphicx}
\usepackage{dcolumn}
\usepackage{bm}
\usepackage{color} 
\usepackage{CJK}
\usepackage{dsfont}

\def\la{{\langle}}
\def\ra{{\rangle}}

\newcommand{\beq}{\begin{equation}}
\newcommand{\eeq}{\end{equation}}
\newcommand{\beqa}{\begin{eqnarray}}
\newcommand{\eeqa}{\end{eqnarray}}

\begin{document}
\title{Fast bias inversion of a double well without residual particle excitation}
\author{S. Mart\'\i nez-Garaot}
\email{sofia.martinez@ehu.eus}
\affiliation{Departamento de Qu\'{\i}mica F\'{\i}sica, UPV/EHU, Apdo
644, 48080 Bilbao, Spain}
\author{M. Palmero}
\affiliation{Departamento de Qu\'{\i}mica F\'{\i}sica, UPV/EHU, Apdo
644, 48080 Bilbao, Spain}
\author{D. Gu\'ery-Odelin}
\affiliation{Laboratoire de Collisions Agr\'egats R\'eactivit\'e, CNRS UMR 5589, IRSAMC, Universit\'e de Toulouse (UPS), 118 Route de Narbonne, 31062 Toulouse CEDEX 4, France}
\author{J. G. Muga}
\affiliation{Departamento de Qu\'{\i}mica F\'{\i}sica, UPV/EHU, Apdo
644, 48080 Bilbao, Spain}
\affiliation{Department of Physics, Shanghai University, 200444
Shanghai, People's Republic of China}
\date{\today}
\begin{abstract}
We design  fast bias inversions of an asymmetric double well so that the lowest states  in each well remain so 
and free from residual motional excitation.
This cannot be done adiabatically, and a sudden bias switch produces 
in general motional excitation. The residual excitation is suppressed by complementing a 
predetermined fast bias change with  a
linear ramp whose  time-dependent slope 
compensates for the displacement of the wells. The process, combined with vibrational multiplexing
and demultiplexing, can produce vibrational state inversion without exciting internal states, just by deforming the trap.    
\end{abstract}
%
%
\maketitle
%
%
%
\section{Introduction}
\begin{figure}[t]
\begin{center}
\hspace*{-.3cm}\includegraphics[width=9.5cm]{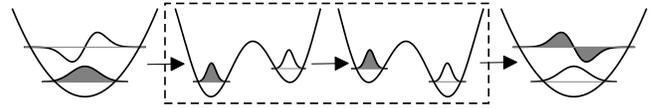}
\end{center}
\caption{\label{general_scheme}
Schematic  representation of demultiplexing (left arrow), bias inversion (framed in dashed line, central arrow) and multiplexing (right arrow).}
\end{figure}

In a recent paper \cite{multi} a protocol to realize fast vibrational state multiplexing/demultiplexing  of ultra cold atoms was introduced.  By a properly designed time-dependent potential deformation between a harmonic trap and a biased double well, the states of a single atom in a harmonic trap can be dynamically mapped into states localized at each well  (vibrational demultiplexing, see the left arrow in Fig. 1), or vice versa (multiplexing, see the right arrow in Fig. 1), faster than adiabatically and without residual  excitation at the final time. 
It was suggested that these processes may be combined with a bias inversion to produce state inversions, from the ground to the first excited state of the harmonic trap and vice versa, 
based on trap deformations only, 
see Fig. \ref{general_scheme}, or to produce vibrationally excited Fock states from an initial ground state \cite{multi}.  
These are basic operations to implement quantum
information processing. Thus the possibility to perform them without exciting internal atomic states as an intermediate step  
is of much interest, as decoherence induced by decay would be suppressed; moreover the systems amenable of manipulation
would not need to have an isolated two-level structure. 
The objective of this paper 
is to design fast controlled bias inversions so that the lowest states 
in each well remain so 
without residual excitation. Unlike multiplexing, 
however, there is no truly adiabatic slow process that achieves this state transformation. In the bias inversion depicted within the central frame of  Fig. \ref{general_scheme},  for example, a slow bias change would preserve the state ordering so that the states represented in the third potential configuration would be interchanged (i.e. the grey state in the right well, and the white one in the left well).     
Nevertheless, in the limit in which the two wells are effectively independent, which in practice means, for times 
shorter than the tunneling time among the wells, the intended state transition might indeed be done slowly enough to be considered adiabatic. If we approximate  each ``isolated'' well by a harmonic oscillator, the intended transformation amounts to a ``horizontal'' displacement along the inter well-axis together with a rising/lowering of the energy of the wells.  The latter effects, however, do not affect the intrawell dynamics, so we may focus on the displacement. In other words, within the stated approximations the bias inversion amounts to the transport of 
a particle in a harmonic potential. Thus, to achieve a fast transition 
without residual excitation we may use shortcuts to adiabaticity designed to perform fast transport \cite{review}.
Specifically we shall use a compensating-force approach \cite{transport_Erik, transport_Mikel}, equivalent to the fast-forward scaling technique \cite{MN}, based on adding  to the potential a linear ramp with time-dependent
slope  to compensate for the effect of the trap motion in the non-inertial frame of the trap. 
We shall compare this approach with a sudden bias switch,  a fast quasi-adiabatic (FAQUAD) 
approach \cite{FAQUAD}, or a smooth polynomial connection without compensation. 
In Sec. \ref{compensating} we introduce the compensating-force approach and Sec. \ref{alme} describes the  alternative methods.
In Sec. \ref{exam} numerical examples are presented with parameters appropriate for trapped ions in 
multisegmented, linear Paul traps, and for neutral atoms in optical traps.  
Finally, in Sec. \ref{discussion} we discuss the results and open questions.
%
%
%
%
%
%
%
%
\section{Compensating-force approach}
\label{compensating}
If the double well potential with nearly independent wells is subjected to a bias inversion such that the trap frequencies of each well are essentially equal and constant
throughout, 
and the trajectories of the well minima move in parallel, the process may be described by a parallel transport 
of two rigid harmonic oscillators, one for each well. 
The Hamiltonian potential near the  minima may be approximated as   
\beq
\label{harmonic_V}
V_{0}(x-x_0)=\frac{1}{2}m\Omega_{0}^2(x-x_{0})^2,
\eeq
where $\Omega_{0}$ is the angular frequency and $x_0=x_0(t)$ is the common notation for either of the two minima.\footnote{We disregard purely time-dependent functions in each well. Differential
phases among the wells depending on these functions can be ignored since the traps are assumed to be independent.} When needed we may distinguish the minima as  $x_{0,\pm}$, with $x_{0,+}>x_{0,-}$. 
The Hamiltonian $H_{0}=p^2/2m+V_{0}(x-x_0)$ has eigenenergies 
$
E_n=\left(n+\frac{1}{2}\right)\hbar \Omega_{0},
$
and well known normalized eigenstates $\phi_n(x-x_0)$, proportional to Hermite polynomials \cite{Schiff}.  

Adding to the Hamiltonian a linear term with an appropriate time-dependent slope 
the non inertial effect of the motion of the well will be compensated in the trap frame \cite{transport_Erik, transport_Mikel}. 
To define the trap frame consider the following  position and momentum displacement unitary operator
\beq
\label{u_t}
\mathcal{U}=e^{ipx_0(t)/\hbar}e^{-im\dot x_0(t)x/\hbar},
\eeq
where the overdot represents a time derivative. 
Starting from the Schr\"odinger equation
\beq
i\hbar\partial_t|\psi\rangle=H_{0}|\psi\rangle,
\eeq
%
the transformed wave function $|\Phi\rangle=\mathcal{U}|\psi\rangle$ obeys
\beq
\label{int_H}
i\hbar\partial_t|\Phi\rangle=\mathcal{U}H_{0}\mathcal{U}^\dag|\Phi\rangle+i\hbar(\partial_t\mathcal{U})\mathcal{U}^\dag|\Phi\rangle=H'_{0}|\Phi\rangle, 
\eeq
where 
the interaction picture (trap frame) 
Hamiltonian is 
\beq
\label{H_trap}
H'_{0}=\frac{p^2}{2m}+V_{0}(x)+mx \ddot x_0+\frac{1}{2}m{\dot x_0}^2,
\eeq
and $V_0(x)=\frac{1}{2}m\Omega_{0}^2x^2$.
The term $m{\dot x_0}^2/2$ only depends on time, it does not affect the dynamics and can be ignored. 
To compensate the motion of the trap, we add $-mx\ddot x_0$ to $H_{0}$. This produces $-m(x+x_0)\ddot{x}_0$ in the trap frame and the resulting potential in that frame is reduced to $V_0(x)$, 
again neglecting purely time-dependent functions.  $V_0(x)$ does not depend  on time, so any stationary state in this trap frame will remain stationary, an excitations are avoided. 
\section{Alternative methods\label{alme}}
In this section we present three simple alternative approaches to perform the bias inversion. 

{\it {Sudden approach.}} In the sudden approach the potential is changed abruptly from the initial  to the final configuration, but  
the state of the system remains unchanged immediately after the potential change (in general it will evolve afterwards). 
If the target state is $\psi_{tar}$ the resulting fidelity is 
\beq
\label{F_sudden}
F_s=|\langle \psi(0)|\psi_{tar}\rangle|.
\eeq

{\it{Fast quasi-adiabatic approach (FAQUAD)}.}
A quasi-adiabatic method to speed up adiabatic processes when there is one control parameter $\lambda(t)$ 
is based on distributing the adiabaticity parameter homogeneously in time \cite{FAQUAD}. 
For instantaneous levels 0 and 1 this means 
\beq
\label{adia}
\hbar \left | \frac{\la \phi_0|\partial_t\phi_1\ra}{E_0-E_1}\right | =c,
\eeq
where the instantaneous eigenstates $\phi_0$, $\phi_1$ and eigenenergies $E_0$, $E_1$ 
depend on time through their dependence on $\lambda$,
and $c$ is constant. By the chain rule this becomes a first order differential equation for $\lambda(t)$, and  
$c$ is set so that the boundary conditions for $\lambda(t)$ at initial, $t=0$,  and final time $t_f$ are satisfied. 
In the transport of a particle with a harmonic oscillator of angular frequency $\Omega_0$ 
centered at $x_0(t)$ we  set $\lambda(t)=x_0(t)$.  
Using the energies and eigenstates of the first two levels of the harmonic oscillator in Eq. (\ref{adia}), the FAQUAD condition 
becomes simply
\beq
\label{adia2}
\frac{m \dot x_0(t)}{\sqrt{2\hbar m \Omega_0}}=c.
\eeq
The solution is the linear connection $x_0(t)=x_0(0)+[x_0(t_f)-x_0(0)]t/t_f$. 
The minimal $t_f$ for which a zero of excitation energy 
appears is \cite{Bowler,FAQUAD} $2\pi/\Omega_0$. 
 
{\it{Polynomial connection without compensation}.}
The final and initial values of the control parameter may as well be smoothly connected 
without applying any compensation, for example using a fifth order polynomial that assures the vanishing 
of first and second derivatives of the parameter at the boundary times.      
\section{Examples\label{exam}}
In the following examples, the potentials and parameters are adapted for a trapped ion in a multisegmented Paul trap, and for 
a neutral atom in a dipole trap.  
\subsection{Trapped ions}
For a trapped ion we consider a simple double well potential of the form
\beq
\label{V_ions}
V(x,t)=\beta x^4+ \alpha x^2+\gamma x,
\eeq
with $\alpha(t)<0$ and $\beta(t)>0$ \cite{Home,Niza,Mainz}.  $\alpha$ and $\beta$ are assumed to be constant 
and $\gamma\equiv\gamma(t)$ time dependent. The bias inversion implies  a change of sign of $\gamma(t)$ from $\gamma_0>0$ to $-\gamma_0$. 

From 
$
\frac{\partial V}{\partial x} =0
$
the condition for the extrema becomes 
\beq
4\beta x^3+2\alpha x+\gamma=0.
\eeq
It is useful to define
\beqa
\label{const_1}
A&=&0,\; B=\frac{2\alpha}{4\beta},\; 
C=\frac{\gamma}{4\beta},
\\
\label{const_2}
Q&=&\frac{A^2-3B}{9},\, 
R=\frac{2A^3-9AB+27C}{54}.
\eeqa
When $R^2<Q^3$ there are two minima and one maximum. With  $\alpha<0$ and $\beta>0$, 
this is satisfied for  
\beq
\label{limits_gamma}
|\gamma|<\left(\frac{2}{3}\right)^{3/2}\sqrt{-\frac{\alpha^3}{\beta}}.
\eeq
The trajectories of the minima are
\label{sol_extrema}
\beqa
\label{ext_1}
x_{0,\pm}=-2\sqrt{Q}\cos\left({\frac{\theta+(1\pm1) \pi}{3}}\right)-\frac{A}{3},
\eeqa
where $\theta=\arccos\left ({\frac{R}{\sqrt{Q^3}}}\right )$, $0\leq \theta \leq \pi$ 
and the roots are taken to be positive.
Up to second order in $\gamma$ they are 
\beqa
\label{trajec_approx_1}
x_{0,-}&\approx&-\frac{1}{\sqrt{2}}\sqrt{-\frac{\alpha}{\beta}}+\frac{\gamma}{4\alpha}-\frac{3\gamma^2\sqrt{-\alpha\beta}}{16\sqrt{2}\alpha^3}, \\
\label{trajec_approx_2}
x_{0,+}&\approx&\frac{1}{\sqrt{2}}\sqrt{-\frac{\alpha}{\beta}}+\frac{\gamma}{4\alpha}+\frac{3\gamma^2\sqrt{-\alpha\beta}}{16\sqrt{2}\alpha^3}.
\eeqa
The quadratic term $\gamma$ is negligible with respect to the linear term when  
\beq
\label{wef_bias}
|\gamma| \ll \frac{4\sqrt{2}}{3}\sqrt{-\frac{\alpha^3}{\beta}},
\eeq
which implies that the trajectories for the minima move in parallel.
Note that this inequality automatically implies the one in Eq. (\ref{limits_gamma}). 
Neglecting the quadratic term, the two minima are given by $x_{0,\pm}=\pm\frac{1}{\sqrt{2}}\sqrt{-\frac{\alpha}{\beta}}+\frac{\gamma}{4\alpha}$.  
The distance between the minima is given by
\beqa
\label{D_minima}
D&=&2\sqrt{Q}\left\{ \cos{ \left( \frac{\theta}{3} \right) }+ \sin{\left [ \frac{1}{6} ( \pi + 2 \theta)\right ]} \right \}
\nonumber\\
%
%
&\approx& \sqrt{2} \sqrt{-\frac{\alpha}{\beta}}+ \frac{3\sqrt{-\alpha \beta}}{8 \sqrt{2} \alpha^3} \gamma^2. 
\eeqa
We may also compute the energy bias between the two wells as
\beq
\label{bias}
\delta=\gamma D.
\eeq
The distance travelled by each well is, when (\ref{wef_bias}) is fulfilled, 
$d=\gamma_0/(2\alpha)$, 
see Eqs. (\ref{trajec_approx_1}) and (\ref{trajec_approx_2}), and
the effective frequency at each minimum  
\beq
\label{w_ef}
\omega_{0}=\sqrt{\frac{1}{m}\left( \frac{d^2V}{dx^2}\right)_{x=x_{0}}}.
\eeq
For Eq. (\ref{V_ions}) the effective frequencies are
\beqa
\label{w_ef_exp}
\omega_{0,\pm}&=&\sqrt{\!\frac{2}{m}} \sqrt{\!\alpha\!+\!\frac{2}{3}\beta \left \{\!A\!+\! 6\sqrt{Q} \cos\!{\left[\frac{1}{3}\! \left (\! \frac{\theta+(1\pm 1)\pi}{3}\! \right)\!\right] }\! \right\}^{\!2}\!}
\nonumber\\
%
%
\label{w_ef2}
&\approx& 2 \sqrt{-\frac{\alpha}{m}} \mp \frac{3}{2\sqrt{2}}\sqrt{\frac{\beta}{\alpha^2 m}} \gamma.
\eeqa
Hence, comparing the two terms, the condition for the frequencies to be essentially constant 
$\omega_{0,\mp}\approx\Omega_0\equiv2 \sqrt{-\frac{\alpha}{m}}$ is   
again the inequality in  Eq. (\ref{wef_bias}). 

\begin{figure}[t]
\begin{center}
\includegraphics[height=4.5cm,angle=0]{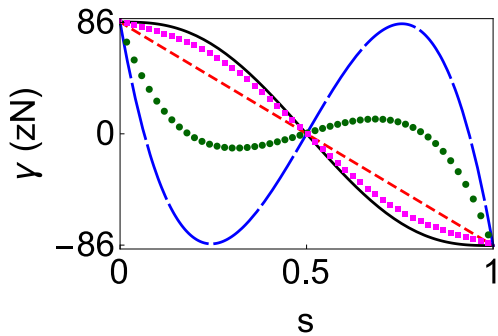}
\end{center}
\caption{\label{displacement_i}
(Color online)
$\gamma$ versus $s=t/t_f$ for the polynomial in Eq. (\ref{connection_i}) (solid black line) and FAQUAD (short-dashed red line). 
$\gamma_0=86.4$ zN, $\gamma(t_f)=-\gamma_0$, $\alpha=-4.7$ pN/m, and $\beta=5.2$ mN/m$^3$.
Also shown are the different effective slopes adding a compensation to the polynomial, 
$\gamma_{eff}(t)=\gamma(t)-m\ddot{\gamma}(t)/(4\alpha)$, 
for the mass of $^{9}$Be$^{+}$ and times $t_f=0.07$ $\mu$s (long-dashed blue line);  $t_f=0.1$ $\mu$s (green dots);  
and $t_f=0.3$ $\mu$s (magenta squares). 
}
\end{figure}

In the regime where the inequality (\ref{wef_bias}) holds, we can apply the compensating force approach to implement a fast bias inversion.
Since the compensating term $-mx\ddot x_0$ is equal for both harmonic traps, we add it to  
$V$ in Eq. (\ref{V_ions}), and the resulting Hamiltonian $H$ is
\beq
\label{H_comp_ions}
H=\frac{p^2}{2m}+\beta x^4+ \alpha x^2+ (\gamma-m\ddot x_0) x.
\eeq
Note that the compensation amounts to changing the time dependence of the slope 
of the linear term from 
the reference process defined by $\gamma(t)$ to $\gamma_{eff}(t)\equiv\gamma(t)-m\ddot x_0(t)=\gamma(t)-m\ddot{\gamma}(t)/(4\alpha)$.  To set $\gamma(t)$ we design a connection between the initial and final configurations.
First note the boundary conditions  
\beq
\label{b_c1}
\gamma (0)=\gamma_0>0, \, \, \, \gamma (t_f)=-\gamma_0,
\eeq
which we complement by 
\beqa
\label{b_c2}
&&\dot \gamma (t_b)=0, \, \, \,  \ddot \gamma (t_b)=0,
\\ \nonumber
&&t_b=0,t_f, 
\eeqa
so that $\dot{x}_0(t_b)=\ddot{x}_0(t_b)=0$. This implies that ${\cal{U}}(t_b)=e^{ip x_0(t_b)/\hbar}$ and  $\dot{\cal{U}}(t_b)=0$. Therefore 
the Hamiltonians and the wave functions in interaction and Schr\"odinger pictures transform into each other by a simple coordinate displacement. 
At intermediate times, we interpolate the function as $\gamma (t)=\sum_{n=0}^{5}c_n t^n$, where the coefficients are found by solving Eqs. (\ref{b_c1}) and (\ref{b_c2}). Finally,
\beqa
\label{connection_i}
\gamma (t)&=&\gamma (0)+ 10 [\gamma (t_f)-\gamma (0)]s^3
\nonumber 
\\ 
&-&15 [\gamma (t_f)-\gamma (0)]s^4+ 6 [\gamma (t_f)-\gamma (0)]s^5,
\eeqa
where $s=t/t_f$. This function and examples of $\gamma_{eff}$ are shown in Fig. \ref{displacement_i}. 
%
%
%
%
%
%
%
%

In order to compare the robustness of the compensating force method against the alternative ones we consider a  $^{9}$Be$^{+}$ ion in the double well with the realistic parameters $\alpha=-4.7$ pN/m and $\beta=5.2$ mN/m$^3$, similar to those in \cite{Wineland}. 
For a moderate initial bias compared to the vibrational quanta, such as 
\beq
\label{moderate_bias}
\gamma_0 \sim \frac{\hbar \Omega_0}{D},
\eeq
the fidelity provided by the sudden approach is one for all practical purposes 
so we can change the bias abruptly and reach the target state.
The  displacement of the trap $d$ may be compared with the oscillator characteristic length 
$a_0=\sqrt{\hbar/m\Omega_0}$. Their ratio is 
\beq
\label{des_rel}
R=\frac{d}{a_0}=\frac{\gamma_0}{2\alpha}\sqrt{\frac{m\Omega_0}{\hbar}}.
\eeq
For the Paul trap $R\approx 0.00065$,  
which explains the high fidelity of the sudden approach for a moderate bias inversion of the ion.   
At these bias values there is really no need to apply a more sophisticated protocol than the sudden switch.    

Henceforth,  we assume a much larger $\gamma_0$, but still satisfying the condition (\ref{wef_bias}).  
In particular, for an initial bias of 1000 $\Omega_0\hbar$ (corresponding to $\gamma_0=86.4$ zN), the ratio becomes $R\approx0.65$.
The maximum variation of the difference between the trajectories of the minima is $3$ pm, about three orders of magnitude less than the displacement of each minimum ($9.2$ nm), so the minima follow parallel trajectories. Furthermore, the maximum variation of the frequencies in Eq. (\ref{w_ef_exp}) with respect to $\Omega_0=2\pi\times 5.6$ MHz
is  $2\pi\times 3.7$ kHz, so the effective frequency is nearly constant.  
%
%
%
%
\begin{figure}[t]
\begin{flushleft}
\includegraphics[width=12.4cm,angle=0]{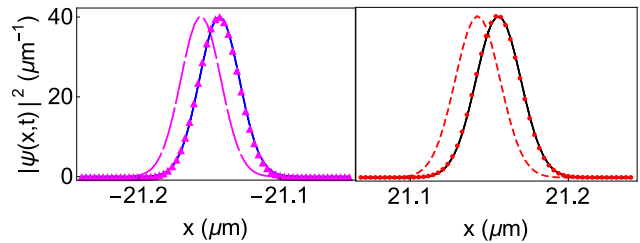}
\end{flushleft}
\caption{\label{cf_evolution_i}
(Color online)
Left: Ground state of the left well  at $t=0$ (long-dashed magenta line) and at $t=t_f$ (magenta triangles), and final state with the compensating force applied on the double well (solid blue line). 
Right: Ground state of the right well: at $t=0$ (short-dashed red line) and at $t=t_f$ (red dots) and final state with the 
compensating force applied (solid black line). 
$t_f=4$ ns and other parameters as in Fig. \ref{displacement_i} for $^{9}$Be$^{+}$.}
\end{figure}
%
%
%
%

Figure \ref{cf_evolution_i} demonstrates  the effect of the 
compensating-force approach. Starting from the ground state of the lower (left) well,  the final evolved state following the shortcut with compensation stays as the ``ground state'' of the left well. This is actually defined as the lowest  state of the double well system predominantly located on the left. 
There is a similar process for the right well. 
The final states represented are obtained by solving the Schr\"odinger equation with the full Hamiltonian (\ref{H_comp_ions}). 
%
%
%
%
\begin{figure}[t]
\begin{center}
\includegraphics[height=2.8cm,angle=0]{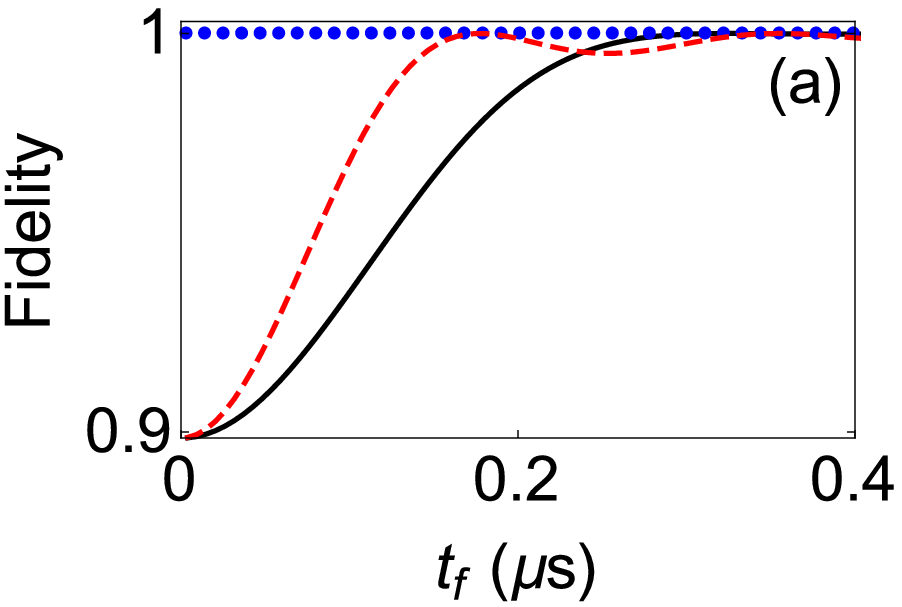}
\includegraphics[height=2.8cm,angle=0]{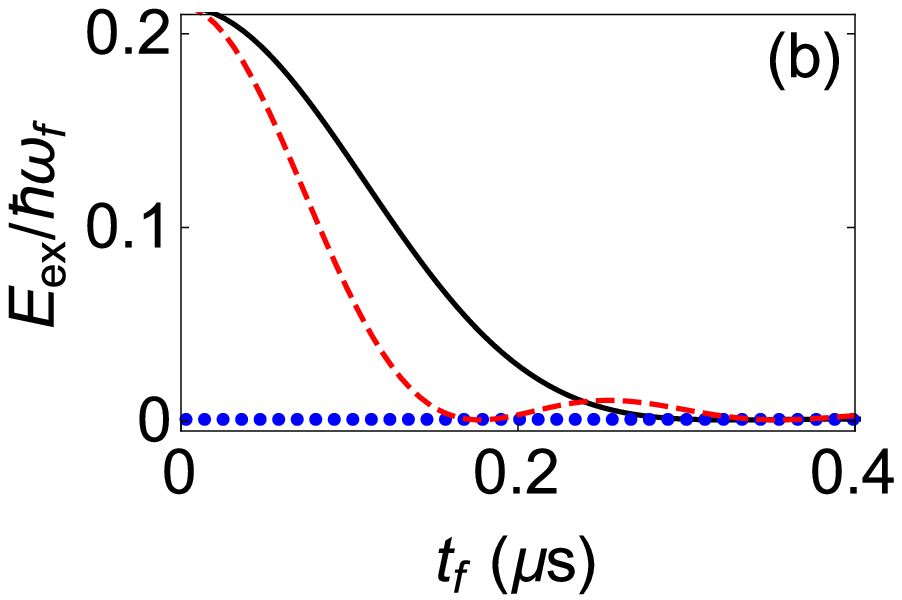}
\end{center}
\caption{\label{dynamics_tf_i}
(Color online).
(a) Fidelity $|\langle\phi_L(t_f)|\psi(t_f) \rangle|$, where $|\phi_L(t_f)\rangle$ is the lowest state located in the 
left well 
in the final time configuration, and $|\psi(t_f) \rangle$ is the evolved state following the shortcut at final time. (b) Final excitation energy for the process on the left well using  
compensating-force (blue dots), fifth degree polynomial in Eq. (\ref{connection_i}) (solid black line), and FAQUAD (short-dashed red line).
The parameters  are for $^{9}$Be$^{+}$ as in Fig. \ref{displacement_i}.}
\end{figure}
%
%
%
%
%
Fig. \ref{dynamics_tf_i} is for the process in the left well. 
It compares the fidelity at final time and the excitation energy, defined as $E_{ex}=E(t_f)-E_0(t_f)$ where $E(t_f)$ is the final energy after the quantum evolution following the shortcut and $E_0(t_f)$ is the ground state final energy of the upper harmonic well,
using the polynomial (\ref{connection_i}) for $\gamma$ 
with and without compensation, as well as the results of the  FAQUAD approach.  
The fidelity without compensation tends to the fidelity of the sudden approach ($0.89$) for very short final times. 
The method with compensation clearly outperforms the others. 
In principle a fundamental limitation of the approach is due to the fact that the inequality (\ref{wef_bias}), that guarantees 
parallel motion and stable frequencies of the wells, should as well be satisfied by $\gamma_{eff}$, but,  at very short times, the dominant 
term of $\gamma_{eff}\sim-m\ddot{\gamma}/{4\alpha}$ may be too large. To estimate this short time limit we 
combine the mean-value theorem inequality for the maximum \cite{transport_Erik}, $|\ddot{\gamma}|_{max}\ge 4\gamma_0/t_f^2$, with Eq. (\ref{wef_bias}) for $\gamma_{eff}$ 
to find  the condition
\beq
 t_f>>\left( \frac{3m\gamma_0}{4\sqrt{2}} \sqrt{-\frac{\beta}{\alpha^5}}  \right)^{1/2}.
\eeq
The factor on the right hand side is 
$10^{-9}$ s for this example, see Fig. \ref{dynamics_tf_i}, which is  several orders of magnitude smaller than $2\pi/\Omega_0$ and does not affect the fidelity in the scale of times shown.  
\subsection{Neutral atoms\label{na}}
The potential
\beq
\label{V_Oberthaler}
V_{na}(x,t)=\frac{1}{2}m\omega^2x^2+V_0\cos^2 \left [ \frac{\pi(x-\Delta x)}{d_l} \right ].
\eeq
forms also a double well. It was implemented in \cite{Oberthaler} with optical dipole potentials, combining a harmonic confinement due to a crossed beam dipole trap with a periodic light shift potential provided by the interference pattern of two mutually coherent laser beams. $\omega$ is the angular frequency of the dipole trap
about the waist position, $V_0$ the amplitude, 
$\Delta x$ the displacement of the optical lattice relative to the center of the harmonic well,  and $d_l$ is the lattice constant. (Double wells with a controllable bias may be also realized by  
two optical lattices of different periodicity with controllable
intensities and relative phase \cite{Bloch}.) Here, the bias inversion implies a change of sign of $\Delta x(t)$ from $\Delta x_0>0$ to $-\Delta x_0$. 

To check if the conditions to apply the compensating force approach hold here 
as well, an analysis similar to the one in the previous example is now performed.
We approximate the potential around each minimum, $V^{(\pm)}$, up to fourth order.  
From $\frac{\partial V^{(\pm)}}{\partial x}=0$ we get a cubic equation for each minimum. 
The positions of the minima are thus given by 
\beq
\label{sol_an}
x_{0,\pm}=-2\sqrt{Q}\cos{\left ( \frac{\theta^{(\pm)}-2\pi}{3} \right )}-\frac{A^{(\pm)}}{3},
\eeq
where
\beqa
Q&=&\frac{2d_l^2\pi^2V_0+d_l^4m\omega^2}{4\pi^4V_0}, 
\nonumber \\
A^{(\pm)}&=&-\frac{3}{2}(2\Delta x \pm d_l), 
\nonumber \\
\theta^{(\pm)}&=&\cos{ \left [ \frac{-3d_l(2\Delta x \pm d_l) m \pi^2\sqrt{V_0}\omega^2}{2(2\pi^2V_0+d_l^2m\omega^2)^{3/2}} \right ]}^{-1}. 
\eeqa
Up to a second order in $\Delta x$,
\beq
\label{sol_an_2}
x_{0,\pm}\approx\pm a+b \Delta x \pm c\Delta x^2, 
\eeq
where the coefficients are known explicitly but too lengthy to be displayed here. 
Whenever the quadratic term is negligible with respect to the linear term ($c\Delta x^2<<b\Delta x$), we can approximate $x_{0,\pm}=\pm a+b\Delta x$ (parallel motion). 
The distance between the minima is given by
\beqa
\label{D_an}
D&=&\frac{1}{3}\left \{ A^{(-)}-A^{(+)}+6\sqrt{Q} \left [ -\cos{\left ( \frac{\pi+\theta^{(-)}}{3} \right )} \right. \right. \nonumber \\ 
&+&\left. \left. \cos{\left (\frac{\pi+\theta^{(+)}}{3} \right )} \right ] \right \}
%
%
%
\approx 2a + 2c\Delta x^2.
\eeqa
Moreover, 
%
$\omega_{0,\pm}\approx f \mp g \Delta x$, again with known but lengthy 
coefficients $g$ and  $f$. 
%
As long as $g\Delta x<<f$
we may set $\omega_{0,\pm}\approx\Omega_0\equiv f$.

For realistic parameters the conditions for parallel transport of the minima and constant frequency are indeed satisfied.  
We consider a ${}^{87}$Rb atom in the trap and
set the parameters in \cite{multi} after the demultiplexing process, namely, $d_l=5.18$ $\mu$m,  $\omega=2\pi\times 59.4$ Hz, and
$V_0/h=1.4$ kHz; the time-dependent displacement  $\Delta x=\Delta x(t)$ is the control parameter
to perform the bias inversion, such that 
\beq
\label{b_c_d0}
\Delta x (0) =\Delta x_0, \, \, \, \Delta x (t_f)=-\Delta x_0,
\eeq
with 
$\Delta x_0=200$ nm. 
We also impose 
\beqa
\label{b_c_d}
&&\dot \Delta x (0)=0, \, \, \,  \ddot \Delta x (0)=0,  \nonumber
\\
&&\dot \Delta x (t_f)=0, \, \, \,  \ddot \Delta x (t_f)=0, 
\eeqa
to achieve similar conditions in the derivatives of the minima $x_0$.
At intermediate times, we interpolate the function as $\Delta x (t)=\sum_{n=0}^{5}d_n t^n$, where the coefficients are found by solving Eqs. (\ref{b_c_d0}) and (\ref{b_c_d}). Consequently, the connection between the initial and final Hamiltonians is given by the same polynomial 
in Eq. (\ref{connection_i}) changing $\gamma(t)\to\Delta x(t)$. 
The double wells are much deeper and tight for trapped ions than for neutral atoms,
compare an intrawell angular frequency $\Omega_0$ of  $2\pi\times5.6$ MHz for the ions versus $2\pi\times0.35$ kHz
for the optical trap. Therefore, in this case, for a moderate initial bias compared to the vibrational quanta, the ratio between the displacement of the trap $d$ and the oscillator characteristic length $a_0$ is $R\approx0.67$.

With the parameters given at time $t=0$, the separation of the minima is $D=5$ $\mu$m, the bias between minima $\delta=2.02\times10^{-32}$ J, and an the effective angular frequency $\Omega_0=2\pi\times 0.35$ kHz, whereas the maximum variation of the frequencies along the process is approximately $2\pi\times 0.2$ Hz. 
Furthermore, the maximum deviation from $D$ of the minima separation is $1.8$ nm, whereas the displacement of each minimum is about $0.4$ $\mu$m. In summary the minima do move in parallel with constant, equal frequencies for practical purposes.  

To accelerate the bias inversion we add the compensating term to $V$ in Eq. (\ref{V_Oberthaler}), 
\beq
\label{H_comp}
H=\frac{p^2}{2m}+\frac{1}{2}m\omega^2x^2+V_0\cos^2 \left [ \frac{\pi(x-\Delta x)}{d_l} \right ]-mx\ddot x_0.
\eeq
%

%
%
%
%
\begin{figure}[h]
\begin{center}
\includegraphics[width=8.6cm,angle=0]{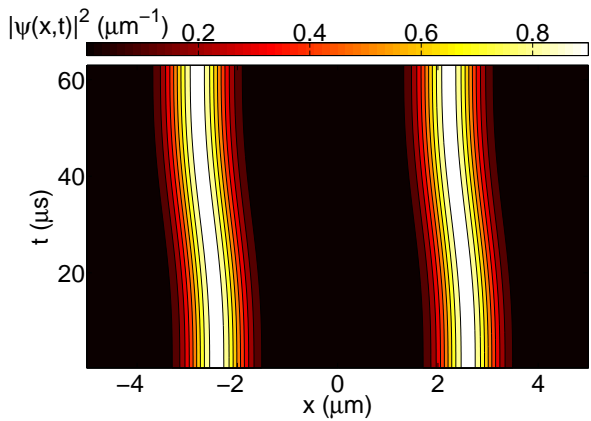}
\end{center}
\caption{\label{contour}
(Color online)
Evolution of the wave function densities  following the shortcut in Eq. (\ref{H_comp}) for 
states in left and right wells.
The parameters are for ${}^{87}$Rb: $d_l=5.18$ $\mu$m, $\omega=59.4\times 2\pi$ Hz, $V_0/h=1.4$ kHz,  $\Delta x_0=200$ nm, and $t_f=63$ $\mu$s.}
\end{figure}
%
%
%
%
Figure \ref{contour} shows the evolution of the densities. 
%
%
%
%
\begin{figure}[t]
\begin{center}
\includegraphics[height=2.8cm,angle=0]{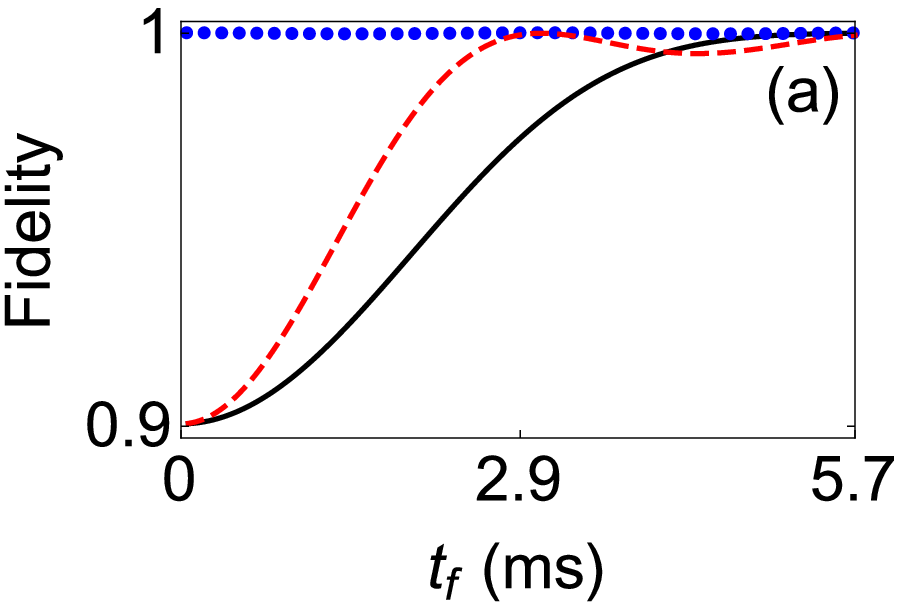}
\includegraphics[height=2.8cm,angle=0]{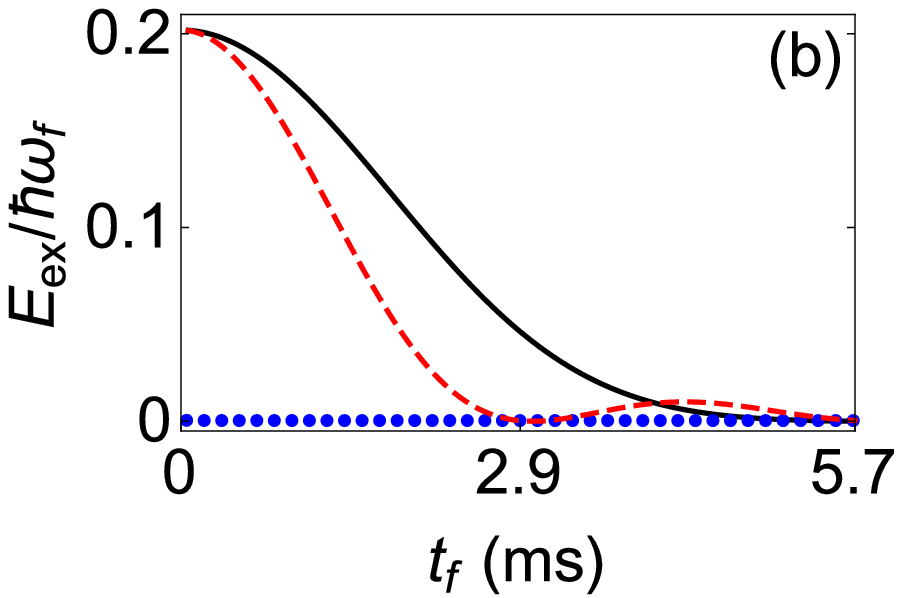}
\end{center}
\caption{\label{dynamics_tf}
(Color online).
(a) Fidelity $|\langle\varphi_1(t_f)|\psi(t_f) \rangle|$, where $|\varphi_1(t_f)\rangle$ is the lowest state 
predominantly of the left well at final time (the first excited state of the double well) and $|\psi(t_f) \rangle$ is the evolved state following the shortcut at final time.
(b) Final excitation energy.
Compensating-force approach (blue dots), fith degree polynomial in Eq. (\ref{connection_i}) with the change $\gamma(t)\to\Delta x(t)$ without compensation (solid black line), and FAQUAD approach (short-dashed red line).
The parameters are chosen for ${}^{87}$Rb: $d_l=5.18$ $\mu$m, $\omega=59.4\times 2\pi$ Hz, $V_0/h=1.4$ kHz, and $\Delta x_0=200$ nm.
}
\end{figure}
%
%
%
%
Focusing on the left well, 
Fig. \ref{dynamics_tf} (a) demonstrates that  the fidelity is exactly one (blue dots) with the compensating force. However, using for the inversion  
the fifth degree polynomial in Eq. (\ref{connection_i}) (with the change $\gamma(t)\to\Delta x(t)$) without compensation the fidelity at short final times decreases until the value of the sudden approach, 0.9.
Furthermore, the excitation (residual) energy $E_{ex}$
is approximately zero for the shortcut protocol, compared to the excitation without compensation 
in Fig. \ref{dynamics_tf} (b). 
In practice, the minimal process times $t_f$ are not limited by the method per se but by the technical capabilities to implement the maximal compensating force. This force depends on the maximal acceleration of the well, whose lower bound is known to be $a_{\rm max}=2d/t^2_f$ \cite{transport_Erik}. 
To implement the compensation with a magnetic field gradient $G$, the gradient should be of the order of $G\simeq m a_{\rm max}/\mu_B$ in an amount of time $t_f$ ($\mu_B$ is Bohr's magneton). For rubidium atoms polarized in the magnetic level $F=m_F=2$ and the transport parameters considered above, this requires a magnetic field gradient on the order of $40$ T/m shaped on a time interval $t_f=63$ $\mu$s. This is definitely challenging from an experimental point of view. Alternatively, one can use the dipole force of an out of axis Gaussian laser beam. If the double well is placed on the edge at a distance of $w/2$ where $w$ is the waist and if $\alpha_p$ denotes the polarizability, the local potential slope experienced by the atoms is on the order of $\alpha_p P/w^3$ where $P$ is the power of the beam. The compensation requires that $P/w^3=m/\alpha_p$. For instance, with an out-of-resonance beam at a wave length of 1 $\mu$m, the polarizability of rubidium-87 atoms is $\alpha_p \simeq 1.3\times 10^{-36}$ m$^2$s, and the compensation can be performed using a 1W laser with a waist of 20 $\mu$m.
\section{Discussion
\label{discussion}}
In this work we have proposed a method to invert the bias of a double-well potential, in the regime of independent wells,  to keep the final states motionally unexcited within the same original well. The method  treats the bias inversion as a rigid transport of the wells, which is justified for realistic parameters, and applies  
a ``compensating-force'' to cancel the excitations. Examples have been worked out 
for ions or neutral atoms and comparisons have been provided with a sudden approach, a fast quasi-adiabatic (FAQUAD) approach, or a smooth polynomial connection of initial and final bias 
without compensation. The compensating-force method clearly outperforms the others and gives ideally perfect fidelities, at least under ideal conditions, up to very small times compared e.g. with the time 
$2\pi/\Omega_0$ where FAQUAD provides a first zero of excitation.   

One more advantage is the flexibility, as the reference process used to design the corresponding compensation
(we have used a polynomial for simplicity) 
may be chosen among a broad family of functions satisfying Eqs. (\ref{b_c1}, \ref{b_c2}). As in other shortcut approaches, this flexibility may be used to enhance robustness versus noise and perturbations \cite{error,Lu,DavidGon}.

The bias inversion put forward here and the multiplexing and demultiplexing protocols developed in \cite{multi} provide  the necessary toolkit to perform vibrational state inversions \cite{Jurgen,Jurgen2}, or Fock state preparations \cite{multi}. Applications in optical waveguide design are also feasible \cite{optical}.  As well, the fast bias inversion 
is directly applicable to Bose-Einstein condensates \cite{Masuda2,transcond}.  

Generalizations for conditions in which rigid transport does not hold 
are also possible using invariant theory \cite{transport_Erik}, which allows for finding processes without final excitation where both the frequency and position of the well depend on time \cite{new}.

\section*{Acknowledgments}
This work was supported by 
the Basque Country Government (Grants No.
IT472-10),
Ministerio de Econom\'{i}a y Competitividad (Grant No.
FIS2012-36673-C03-01), and the program UFI 11/55.
S. M.-G. and M. P. acknowledge fellowships by UPV/EHU.  
%
%
%
%


\end{document}